# Impact of Weather Conditions on 5G Communication Channel under Connected Vehicles Framework

Esmail M M Abuhdima, Ahmed El Qaouaq, Shakendra Alston, Kirk Ambrose, Gurcan Comert, Jian Liu, Chunheng Zhao, Chin-Tser Huang, and Pierluigi Pisu

*Abstract*— Recent research focused on improving the vehicle-to-vehicle communication (V2V) based on the 5G technology. The V2V application is important because it will reduce the risk of accidents up to 70%-80%, improve traffic management, reduce congestion, and improve fuel consumption. Autonomous vehicles applications require a high bandwidth transmission channel where the 5G communication channel would be a reliable solution to support this disruptive technology. The dedicated short-range communications (DSRC), which is characterized with a frequency bandwidth of 5.9 gigahertz (GHz) (4G spectrum), was used as vehicular connectivity with a bandwidth of up to 200 megabytes per second (mb/s) and limited capacity. The 5G band can support connected multiple autonomous vehicles with high data rates and large bandwidth. In this study, the 5G communication channel is considered as vehicular connectivity with high bandwidth in the millimeter waves spectrum range. The quality of 5G wireless communication channels between connected vehicles possibly be affected by weather conditions such as rain, snow, fog, dust, and sand. In this paper, we estimate the effect of dust and sand on the propagation of millimeter waves. The Mie model is used to investigate the effect of dust and sand storms on the propagating mm-waves. The effect of dust and sand on the communication path loss of DSRC and 5G frequency band is investigated in the case of urban freeway and rural highway settings. Results show that the attenuation of dust and sand is changed when the particle size of sand, frequency of propagating wave, and concentration of dust are changed. Finally, the new model of link margin is created to estimate the effect of dust and sand on DSCR (5.9 GHz) and 5G (28 GHz) communication path loss.

## I. INTRODUCTION

INTELLIGENT transportation systems (ITS) applications have been shaped by upcoming technological innovations in order to improve safety, mobility, efficiency, and decrease adverse impacts. Advancements in communication technologies enable different applications such as vehicle-to-vehicle (V2V), vehicle-to-infrastructure (V2I), vehicle-to-pedestrian (V2P), and vehicle-to-network (V2N). Real-time data sharing within these applications is critical as they help increase traffic efficiency, reduce crashes, ease traffic congestion, and improve fuel efficiency [1], [2]. However, as reliance on these applications grows, reliability needs also to increase to prevent any unfortunate incidents. One of the critical adverse elements in sensing and communication technologies is the weather. In this paper, the effect of dust and sands on the propagating mm-Wave will be discussed. The difference between the dust and sand particle is simply the diameter of the sand diameter larger than the dust particle diameter. The dust particle size is ranged from 10 micrometers (μm) to 80μm, and the sand particle size is ranged from 0.15 millimeters (mm) to 0.3 mm. The attenuation caused by the dust and sand is one of the major problems in the use of terrestrial and space wireless communication (Microwave and Millimeter-wave signals). Regarding the fact, one misconnection can cause big damage to circuit traffic. This research is very important to provide a better reliable connection between the vehicles in the existence of this weather condition.

There are limited studies in the literature on the impact of weather on 5G (mm-Wave) communications [3]. Only a

This study was based upon a project funded by the US Department of Transportation (USDOT) Center for Connected Multimodal Mobility (C2M2) (Tier 1 University Transportation Center) headquartered at Clemson University, Clemson, South Carolina, USA and U.S. National Science Foundation Grants No. 1719501 and 1954532.

Esmail Abuhdima with the Dept. of Com. Sci., Phy. and Eng., Benedict College, Columbia, SC USA (esmail.abuhdima@benedict.edu).
Ahmed El Qaouaq with the Dept. of Com. Sci., Phy. and Eng., Benedict College, Columbia, SC USA (ahmed.el-qaouaq34@my.benedict.edu).
Shakendra Alston with the Dept. of Com. Sci., Phy. and Eng., Benedict College, Columbia, SC USA ( shakendra.alston68@my.benedict.edu).
Kirk Ambrose with the Dept. of Com. Sci., Phy. and Eng., Benedict College, Columbia, SC USA (Kirk.Ambrose61@my.benedict.edu).
Gurcan Comert with the Dept. of Com. Sci., Phy. and Eng., Benedict College, Columbia, SC USA (gurcan.comert@benedict.edu).
Jian Liu with the Department of Computer Science and Engineering, University of South Carolina, Columbia, SC USA (jianl@email.sc.edu).
Chunheng Zhao, Int. Center for Automotive Research, Clemson University, Greenville, USA (chunhez@clemson.edu).
Chin-Tser Huang with the Dept. of Computer Science and Engineering, University of South Carolina, Columbia, SC USA (huangct@cse.sc.edu).
Pierluigi Pisu, Int. Center for Automotive Research, Clemson University, Greenville, USA (pisup@clemson.edu).

handful of them discusses the effect on vehicular communications. Previous research studied attenuation and backscattering effects from formulations and quantified the weather particle sizes impacts [1]. In the background literature, the empirical path loss models were developed using measurement data for different environmental areas [4]. These researches considered the effect of distances between vehicles, small and large-vehicle obstruction, climate factors such as rain and fog, and the height and characteristic of the antenna on the path loss. Also, the effect of sand and dust storms on wireless communication, such as microwave links and GSM signal coverage, is addressed, and found that the propagating signal is attenuated by the effect of sand and dust [5] and [6]. In this research, 5G mm-waves are proposed as a solution to the vehicle's connectivity since the 5G can support connected autonomous vehicles with high data rates and huge transmitting bandwidth. For these main reasons, the 5G wireless communication channel is considered as the vehicular connectivity preferable option. This project will consider most of the wireless interactions between the self-driving vehicles on a low-height level, approximately a range between 1m to 4m.

In this paper, the effect of dust and sand with standards of the visibility, particle size, and different humidity levels (0%, 60%, 100%) on the 5G vehicle channel is investigated in comparison with the dedicated short range communication channel, DSRC, (5.9 GHz). Two possible scenarios are considered to estimate the path loss of the propagating signal. These scenarios are urban and highway conditions. These scenarios are according to the main driving categories, but they have different conditions. The path loss for each condition should be calculated differently to estimate the accurate value of the link margin. The specific formula that is used for each condition to estimate the path loss will be shown in Section III. Along with transmission parameters of used sensors such as transmitted power, antenna gain, and total free space loss, the attenuation of the transmitted signal is computed in terms of operating frequency, the concentration of dust, and particle size of sand. According to the received signal, it can determine whether the attenuation of the signal is high or low in comparison with the threshold of the received power. The worst weather condition is considered to calculate the attenuation factor in $dB/km$, when the visibility is low, and the particle size of dust and sand is large.

This paper is separated into five main sections. Section II will generally discuss the main topic of the research that studied the effect of dust and sand storms on the propagating 5G mm-wave. Section III describes the effect of dust and sand on the Link Budget of the communication channel. The region of study and research measurements are mentioned in Section IV. The final results are discussed in Section V. Finally, the conclusion and recommendations are represented in section VI.

## II. SAND AND DUST ATTENUATION

### A. Definition of dust and sand

In general, attenuation is the reduction of propagating signal strength that is derived in terms of visibility, wavelength, and permittivity of transmission media. The deserts across the planet are a resource of the dust and sand particles that are distributed in the world, and those deserts make up approximately 20% of the Earth's surface. For example, the desert of North Africa is the source of dust that affects southern Europe [7]. Also, the desert of the southwest of the United States is the source of dust and sand [7].

The proposed model of dust and sand attenuation $A_d$, can be used to determine the effect of dust and sand storms on the propagating mm-wave when the concentration of the dust and the radius of particle size vary according to change of the weather condition. This attenuation factor is dependent on different main variables or parameters, such as operating frequency, humidity, height, particle size, and visibility. These factors contribute to a variation of the attenuation factor, so if these parameters change, the attenuation factor will also vary. Frequency is the number of occurrences that a wave surpasses a point within a specific amount of time. Frequency is measured in the unit of Hertz. Humidity is the measure of how much water vapor is in the air. Height is a measure of distance above the ground where the attenuation factor is computed. Visibility is the measure of distance that a human being can see an object during dust and sand storms. All of these factors are taken into consideration when the attenuation factor is investigated. The visibility at reference height $h_o$ and reference visibility $V_0$ is defined as in [8] and given in Eq. 1.

$$V^{\gamma} = V_0^{\gamma} \left[ \frac{h}{h_0} \right]^b \quad (1)$$

where $\gamma$ is a constant that depends on the distance from the point of origin of the storm type of soil and climatic conditions at the origin, b is a constant that depends on the climatic conditions, meteorological factors, and the particle size distribution of the dust and the sand, and $h$ is the height from the ground [9].

The important direct measurement to investigate the effect of dust and sand storms is to predict the dielectric constant for dust and sand particles from their mineral and /or chemical composition values. The complex permittivity of the composite component $(\varepsilon_m)$ is computed using the Looyenga equation as given by [10] and presented in Eq. 2.

$$\varepsilon_m^{\frac{1}{3}} = \sum_{i=1}^{n} v_i \varepsilon_i^{\frac{1}{3}} \quad (2)$$

where $\varepsilon_i$ is the complex dielectric constant of the $i^{th}$

substance and $v_i$ is the relative volume of the $i^{th}$ sample from the volume of the total sample. The permittivity of transmission media is written as

$$\varepsilon = \varepsilon_1 - \varepsilon_2 \quad (3)$$

where $\varepsilon_1$ is the dielectric constant, and $\varepsilon_2$ is the dielectric loss factor.

*B. Attenuation Model*

The mathematical model which is based on Mie scattering method, is created to compute the reduction of the propagating mm-wave strength. The ratio of diameter (sand /dust) to the wavelength of the propagating signal is considered when the attenuation factor model is developed to get the accurate effect. This model is valid to use, especially at higher frequencies. The most important parameters that affect the attenuation value are particle radius, operating frequency, humidity, and complex permittivity. The attenuation of the dust and sand $A_d$, is defined by [11] and [12] and presented in Eq. 4.

$$A_d = \frac{a_e f}{v}\left[C_1 + C_2 \, a_e^2 \, f^2 + C_3 \, a_e^3 \, f^3\right] \quad \frac{dB}{km} \quad (4)$$

where

$$C_1 = \frac{6\varepsilon_2}{(\varepsilon_1+2)^2 + \varepsilon_2^2}$$

$$C_2 = \varepsilon_2\left[\frac{6}{5}\frac{7\varepsilon_1^2 + 7\varepsilon_2^2 + 4\varepsilon_1 - 20}{\left[(\varepsilon_1+2)^2 + \varepsilon_2^2\right]^2} + \frac{1}{15} + \frac{5}{3\left[(2\varepsilon_1+3)^2 + 4\varepsilon_2^2\right]}\right]$$

$$C_3 = \frac{4}{3}\left[\frac{(\varepsilon_1-1)^2(\varepsilon_1+2) + \left[2(\varepsilon_1-1)(\varepsilon_1+2)-9\right] + \varepsilon_2^4}{\left[(\varepsilon_1+2)^2 + \varepsilon_2^2\right]^2}\right]$$

$a_e$ is the equivalent particle radius in meters, $v$ is the visibility in kilometer, f is the frequency in GHz, $C = 2.3 \times 10^{-5}$, $\gamma = 1.07$, $g = 1.07$ and $b = 0.28$ [13].

III. LINK BUDGET ANALYSIS WITH DUST AND SAND

First of all, the path loss is the loss or attenuation of the power of a propagating wave as it propagates through transmission media. In this study, the free space is considered as a transmission media. From Eq. 4, when the dust and sand particles in the air increase, the path loss is also increased. Path loss is important because it is needed to find the total received power of signals transmitted from point to point, such as vehicle to vehicle or vehicle to infrastructure. In the case of the V2V connection, there are two possible scenarios that are studied for path loss. These scenarios are urban and highway conditions. These two scenarios were specifically chosen because they are general driving environments regardless of the general locations. The path loss for each case should be calculated differently to commute an accurate value of the signal loss. The path loss of the urban condition is defined in [14]

$$L_s = 38.77 + 16.7\log_{10}(d) + 18.2\log_{10}(f) + \chi_a \quad (5)$$

and the path loss of the Highway condition can write as [14]

$$L_s = 23.4 + 20\log_{10}(d) + 20\log_{10}(f) + \chi_a \quad (6)$$

where $f$ is the signal frequency in GHz, $d$ is the distance between vehicles in meters, and $\chi_a$ represents the shadowing. According to dust and sand storms conditions, the expression of the path loss for both conditions will be modified to take into account the effect of dust and sand. The modified path loss of Urban condition is

$$L_{sm} = 38.77 + 16.7\log_{10}(d) + 18.2\log_{10}(f) + A_d + \chi_a \quad (7)$$

and the modified Highway condition is

$$L_{sm} = 23.4 + 20\log_{10}(d) + 20\log_{10}(f) + A_d + \chi_a \quad (8)$$

where $A_d$ is the attenuation factor of dust and sand that is defined by (4).

The proposed link margin $(M_d)$ is defined as

$$M_d = \frac{G_t G_r P_t}{k T_s R L_0 L_{sm}\left(\dfrac{E_b}{N_0}\right)} \quad (9)$$

where $G_r$ is the receive antenna gain, $G_t$ is transmit antenna gain, $P_t$ is the transmitted power (dBm), $R$ is the data rate $(bps)$, $L_0$ is the circuit losses, $k$ is the Boltzmann's constant $(1.38 \times 10^{-23} \, J/K)$, $T_s$ is the effective system noise temperature, $\dfrac{E_b}{N_0}$ is the required energy per bit to noise spectral density ratio, $L_{sm}$ is the modified free space path loss that is represented by (7) and (8).

IV. REGION OF STUDY

The desert of the southern part of Libya was investigated in [5] for the impact of dust and/or sand storms on the wireless communication systems. This region is considered because it is characterized as a famous desert climate, and it has a fast wind filled with dust from time to time. Based on the study region information, nine places were chosen to collect the sand and dust carried by the wind. The plastic cans were placed on the tower or on the roof of buildings at the height of 13m to collect dust and sand through the dust storms of different seasons summer, fall, winter, and spring.

*A. Analysis of the samples*

For this study, samples' particle size distribution, average density, and chemical composition are required to compute the dielectric constant and attenuation factors. There are two laboratories available to carry out the analysis of the samples. These two laboratories are the Libyan Petroleum Institute and the Industrial research. The measured density of all samples which the Libyan Petroleum Institute has determined is shown in Table I.

Table I. Density of Samples

| Sample No. | Density $(g/cm^3)$ |
|---|---|
| 1 | 2.5426 |
| 2 | 2.56857 |
| 3 | 2.6138 |
| 4 | 2.62714 |
| 5 | 2.4202 |
| 6 | 2.9232 |
| 7 | 2.4732 |
| 8 | 2.5425 |
| 9 | 2.4764 |

The average density of all samples from Table I is equal to 2.5764 g/m3.

The size of dust and sand particles is determined using the sieving method. The sieving analysis showed that the major grain size of these samples ranges from less than 90 μm to 600 μm. The equation (2) is used to calculate the complex permittivity of the composite component. The Complex Permittivity of each sample is shown in Table II.

Table II. Complex Permittivity of each Sample

| Sample N0 | Complex Permittivity |
|---|---|
| 1 | 5.0384 - j 0.0509 |
| 2 | 5.4851 - j 0.0562 |
| 3 | 5.4801 - j 0.0694 |
| 4 | 7.5929 - j 0.1140 |
| 5 | 6.7899 - j 0.1296 |
| 6 | 5.4003 - j 0.0787 |
| 7 | 7.4707 - j 0.1344 |
| 8 | 5.5713 - j 0.0704 |
| 9 | 8.3078 - j 0.1329 |

The average complex permittivity of the samples collected in the studied region from table II is equal to 6.3485 - j 0.0929. The following empirical relation is used for the variation of complex permittivity with relative humidity from [5],

$$\varepsilon_1 = 6.3485 + 0.04H - 7.78 \times 10^{-4} H^2 + 5.56 \times 10^{-6} H^3 \quad (10)$$

and

$$\varepsilon_2 = 0.0929 + 0.02H - 3.71 \times 10^{-4} H^2 + 2.76 \times 10^{-6} H^3 \quad (11)$$

where $H$ is the air relative humidity (percentage).

## V. RESULTS AND DISCUSSION

### A. Attenuation of dust and sand

Equation (4) is used to estimate the attenuation of dust and sand in terms of concentration of dust, the particle size of sand, and frequency of the transmitted signal. The 5G (28GHz) and DSCR (5.9 GHz) two bands of frequency are considered. The particle size is changed between 0 to $538\ um$. The maximum height is between $0.75\ m - 2m$. The attenuation is computed at different values of humidity $(0\%, 60\%, and\ 100\%)$. Fig. 1 shows the attenuation of dust and sand versus the visibility when the transmitted frequency is 5.9GHz.

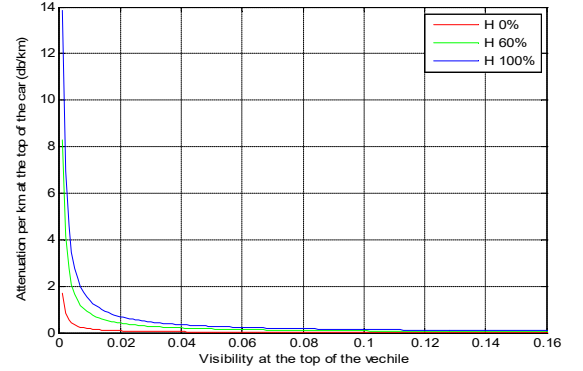

Fig. 1 Attenuation versus the visibility at 5.9GHz and *h*=1 *m*)

The attenuation of dust and sand is increased as the frequency increases, as shown in Fig. 2. Also, the same attenuation model is used to investigate the effect when the particle size is increased at different values of considered frequency. It is found that the value of the attenuation of 5G is higher than the value of DSRC, as shown in Figs. 3 and 4.

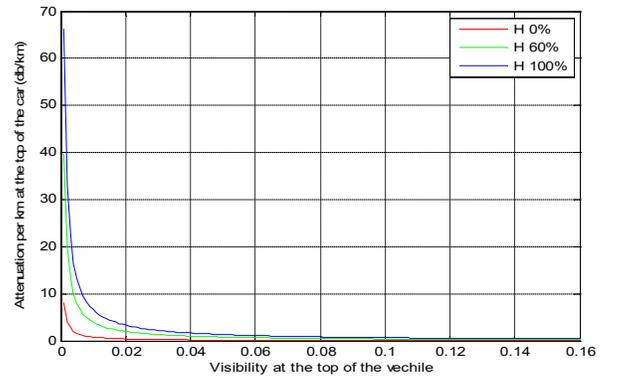

Fig. 2 Attenuation versus the visibility at 28GHz and *h*=1 *m*

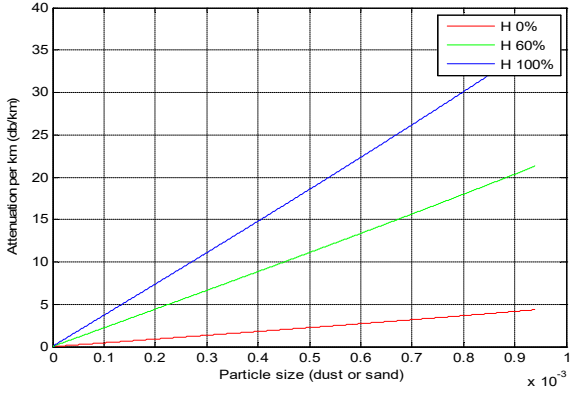
Fig. 3 Attenuation vs. particle size at 5.9GHz and $h=1\ m$

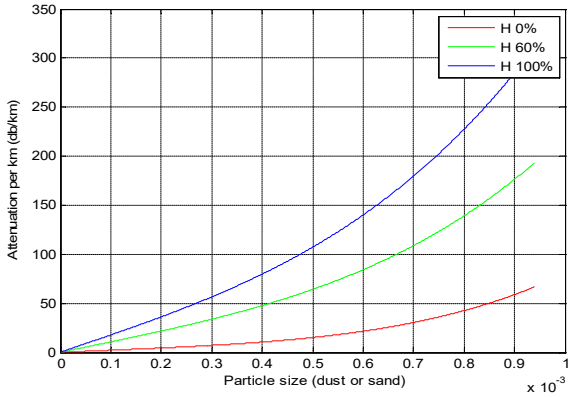
Fig. 4 Attenuation vs. particle size at 28GHz and $h=1\ m$

### B. Link Budget of V2V

The designed system consists of two vehicles. The sensor of every vehicle used a dual-band phased array antenna. This system is proposed and tested by a research team from Florida International University [15]. This dual-band RF front-end for DSRC and 5G V2V is operating at both 5.9GHz and 28GHz. The modified link budget of the V2V communication link that is presented (9) is used to estimate the received power signal. If the received energy per bit to noise spectral density ratio is achieving the required bit error rate, the communication between V2V is reliable. On the contrary, if the received energy per bit to noise spectral density ratio is not achieving the required bit error rate, the communication between V2V is disconnected.

Link Budget information of 5.9 GHz and 28GHz is shown in Table III [15].

Table III. Required link budget information

|  | DSRC 5.9GHz | mm-wave 28GHz |
|---|---|---|
| Transmit Antenna Gain | 9.9 dBi | 23.4 dBi |
| Receive Antenna Gain | 9.9 dBi | 23.4 dBi |
| Transmit power | 27 dBm | 27 dBm |
| EIRP | 36.9 dBm | 50.4 dBm |
| Required $E_b/N_0$ | 18.8 dB | 18.8dB |
| Data Rate | 27 Mbps | 1 Gbps |

According to the proposed model of Sandhiya Reddy Govindarajulu and Elias A. Alwan [15], it assumed the overall circuit losses $L_0 = 5\,dB$, receiver front-end noise figure $F = 6\,dB$, $T_s = T_A + (F-1)290$, $T_A = 290K$, the distance between vehicles $d = 390\,m$ and $\frac{E_b}{N_0} = 18.8\,dB$ that is required to achieve a bit error rate $(BER)$ equal to $10^{-6}$. Moreover, the radio frequency parameters that are presented in Table III are used together to insert into (9) to calculate the value $L_{sm}$ to maintain the link margin $(M_d \geq 10\,dB)$. When $M_d \geq 10\,dB$, all the required transmission parameters are achieved to avoid any disconnect or loss of communication between autonomous vehicles. Since the attenuation $L_{sm}$ at $M_d = 10\,dB$ (threshold value) is computed, Figs. 1, 2, 3, and 4 are used to figure out the minimum equivalent particle radius and visibility (concentration of dust) that avoid the loss communication between V2V. In other words, these minimum values of equivalent particle radius and visibility are the threshold value. If equivalent particle size (P.S) is increased or the visibility is decreased, the communication between V2V will be disrupted. For this reason, this research is important to design a reliable system to support the recent intelligent transportation network. The result of this simulation is summarized in Table IV.

Table IV. Simulation result

| Scenario | | Urban | Highway |
|---|---|---|---|
| 0% | 5.9GHz | V is not effect<br>P.S >1.48 mm | V is not effect<br>P.S > 614 um |
| | 28GHz | V is not effect<br>P.S > 0.27 mm | V < 51 m<br>P.S > 2um |
| 60% | 5.9GHz | V is not effect<br>P.S > 0.328 mm | V < 2 m<br>P.S > 129 um |
| | 28GHz | V < 3m<br>P.S > 58 um | V < 120m<br>P.S > 0.39 um |
| 100% | 5.9GHz | V < 1 m<br>P.S> 0.2 mm | V < 3 m<br>P.S > 70 um |
| | 28GHz | V < 5m<br>P.S > 36 um | V < 160 m<br>P.S > 0.2 um |

Table IV shows that dust and sand affect the 5G communication channel more than the DSRC channel. This effect is logical because the wavelength $(\lambda)$ of the propagating 5G mm-wave is short in comparison with the particle size of dust and sand. It is seen that the 5.9GHz is not affected by the dust in the dry weather (H=0%) for both scenarios, but the 5G (28GHz) is affected when visibility is less than $51\,m$ in the highway case. At the H=0%, the 5.9GHz is affected if the particle size is greater than $1.48\,mm$ in the urban case and $614\,um$ in the highway case. At the H=0%, the 28GHz is affected if the particle size is greater than $0.27\,mm$ in the urban case and $2\,um$ in the highway case.

At the H=60%, the 5.9GHz is affected if the particle size is greater than $0.328\,mm$ in the urban case and $129\,um$ in the highway case, but it is affected by the concentration of dust if the visibility is less than $2\,m$ in the highway case. At the H=60%, the 28GHz is affected if the particle size is greater than $58\,um$ and the visibility less than $3\,m$ in the urban case. It is affected when the particle size is greater and $0.39\,um$ the visibility less than $120\,m$ in the highway case.

At the H=100%, the 5.9GHz is affected if the particle size is greater than $0.2\,mm$ in the urban case and $70\,um$ in the highway case. Also, it is affected by the concentration of dust if the visibility is less than $1\,m$ in the urban scenario and $3\,m$ in the highway scenario. At the H=100%, the 28GHz is affected if particle size is greater than $36\,um$ and the visibility less than $5\,m$ in the urban case. Also, It is affected when the particle size is greater than $0.2\,um$ and the visibility less than $160\,m$ in the highway case.

## VI. Conclusions

In this research, the effect of dust and sand storms on the 5G wireless communication channel between connected vehicles was investigated. The proposed margin link model is used to simulate the effect of dust and sand. The main object point of this work is to find the threshold value of particle size of sand and the concentration to avoid the loss of communication between autonomous vehicles. The simulation results show that the attenuation of the 5G propagating signal increases when the operating frequency, concentration of the dust, and particle size of sand are increased. Also, it is seen that the 5G mm-wave communication channel is more affected by dust and sand storms than the DSRC channel. As future work, this proposed model will be validated by using 5G toolbox or real measurements. Moreover, the real measurement of dielectric constant, particle size range, and concentration of dust of the desert regions should be considered to figure out the values of real attenuation of dust and sand for this specific region.

## References

[1] S. Zang, M. Ding, D. Smith, P. Tyler, T. Rakotoarivelo and M. A. Kaafar, "The Impact of Adverse Weather Conditions on Autonomous Vehicles: How Rain, Snow, Fog, and Hail Affect the Performance of a Self-Driving Car," in IEEE Vehicular Technology Magazine, vol. 14, no. 2, pp. 103-111, June 2019.
[2] R. He, A. F. Molisch, F. Tufvesson, Z. Zhong, B. Ai and T. Zhang, "Vehicle-to-Vehicle Propagation Models With Large Vehicle Obstructions," in IEEE Transactions on Intelligent Transportation Systems, vol. 15, no. 5, pp. 2237-2248, Oct. 2014.
[3] al-saman, Ahmed & Cheffena, Michael & Mohamed, Marshed & Bin Azmi, Marwan & Ai, Yun, "Statistical Analysis of Rain at Millimeter Waves in Tropical Area," IEEE Access. PP. 1-1. 10.1109/ACCESS.2020.2979683.
[4] M. Giordani, T. Shimizu, A. Zanella, T. Higuchi, O. Altintas and M. Zorzi, "Path Loss Models for V2V mmWave Communication: Performance Evaluation and Open Challenges," 2019 IEEE 2nd Connected and Automated Vehicles Symposium (CAVS), Honolulu, HI, USA, 2019, pp. 1-5.
[5] E. M. Abuhdima and I. M. Saleh, "Effect of sand and dust storms on microwave propagation signals in southern Libya," Melecon 2010 - 2010 15th IEEE Mediterranean Electrotechnical Conference, Valletta, 2010, pp. 695-698.
[6] E. M. Abuhdima and I. M. Saleh, "Effect of sand and dust storms on GSM coverage signal in southern Libya," 2010 International Conference on Electronic Devices, Systems and Applications, Kuala Lumpur, 2010, pp. 264-268.
[7] UN, ESCAP, "Global Alarm: Dust and Sandstorms from the World's Drylands," UNCCD, June 2002.
[8] I. M. Saleh, H. M. Abufares and H. M. Snousi, "Estimation of wave attenuation due to dust and sand storms in southern Libya using Mie model," WAMICON 2012 IEEE Wireless & Microwave Technology Conference, Cocoa Beach, FL, 2012, pp. 1-5.
[9] S. Ghobrial and S. Sharief, "Microwave attenuation and cross polarization in dust storms," in IEEE Transactions on Antennas and Propagation, vol. 35, no. 4, pp. 418-425, April 1987.
[10] S.I Ghobrial and S.M Sharief, "X-Band Measurement of the Dielectric Constant of Dust, " Proc.Ursi commission F 1983 Symposium. Louvain .Belgium .June 1983, PP143-147.
[11] Sami M. Sharif, "Attenuation Properties of Dusty Media Using Mie Scattering Solution," Progress In Electromagnetics Research M, Vol. 43, 9–18, 2015.
[12] A. Musa and B. S. Paul, "Prediction of electromagnetic wave attenuation in dust storms using Mie scattering," 2017 IEEE AFRICON, Cape Town, 2017, pp. 603-608.
[13] D Gillett, C Morales, "Environmental Factors Affecting Dust Emission by Wind Erosion," Saharan dust, 1979.
[14] M. Giordani, T. Shimizu, A. Zanella, T. Higuchi, O. Altintas and M. Zorzi, "Path Loss Models for V2V mmWave Communication: Performance Evaluation and Open Challenges," 2019 IEEE 2nd Connected and Automated Vehicles Symposium (CAVS), Honolulu, HI, USA, 2019, pp. 1-5.
[15] S. R. Govindarajulu and E. A. Alwan, "Range Optimization for DSRC and 5G Millimeter-Wave Vehicle-to-Vehicle Communication Link," 2019 International Workshop on Antenna Technology (iWAT), Miami, FL, USA, 2019, pp. 228-230.